\documentstyle[aps,pra,twocolumn,floats,epsfig,curves]{revtex}

\newlength{\absskip}

\newcommand{\ket}[1]{\bigl|#1\bigr\rangle}
\newcommand{\bra}[1]{\bigl\langle #1\bigr|}
\newcommand{\ketbra}[2]{\bigl|#1\bigr\rangle\bigl\langle #2\bigr|}
\newcommand{\proj}[1]{\ketbra{#1}{#1}}
\newcommand{\braket}[2]{\bigl\langle #1\bigr|#2\bigr\rangle}
\newcommand{\Exp}[1]{{\rm e}^{\mbox{\footnotesize$#1$}}}
\newcommand{\adj}{^{\dagger}}\newcommand{\padj}{^{\phantom{\dagger}}}
\newcommand{\repr}{\widehat{=}}
\newcommand{\id}{\openone}
\newcommand{\idtau}{\openone_{\tau}}
\newcommand{\idsig}{\openone_{\sigma}}
\renewcommand{\L}{{\sf L}}
\newcommand{\R}{{\sf R}}
\renewcommand{\v}{{\sf v}}
\newcommand{\h}{{\sf h}}
\newcommand{\UBS}{U_{\rm BS}}
\newcommand{\Umirr}{U_{\rm mirr}}
\newcommand{\UR}{U_{\rm R}}
\newcommand{\UL}{U_{\rm L}}
\newcommand{\UMZ}{U_{\rm MZ}}
\newcommand{\clUMZ}{{\cal U}_{\rm MZ}}
\newcommand{\UQWP}{U_{\rm QWP}}
\newcommand{\clUQWP}{{\cal U}_{\rm QWP}}
\newcommand{\UHWP}{U_{\rm HWP}}
\newcommand{\clUHWP}{{\cal U}_{\rm HWP}}
\newcommand{\Upol}{U_{\rm pol}}
\newcommand{\VR}{V_{\rm R}}
\newcommand{\VL}{V_{\rm L}}
\newcommand{\SRR}{S_{\rm RR}}
\newcommand{\SLL}{S_{\rm LL}}
\newcommand{\SRL}{S_{\rm RL}}
\newcommand{\SLR}{S_{\rm LR}}
\newcommand{\bpsi}{\overline{\psi}}
\newcommand{\bchi}{\overline{\chi}}
\newcommand{\stdbaskets}{\Bigl(%
\ket{\R\v},\ket{\R\h},\ket{\L\v},\ket{\L\h}\Bigr)}
\newcommand{\kBell}[1]{\ket{{\sc b}_{#1}}}
\newcommand{\bBell}[1]{\bra{{\sc b}_{#1}}}

\newcommand{\bVAA}[1]{\bra{{\sc vaa}_{#1}}}

\begin{document}
\draft
\title{Universal unitary gate for single-photon 2-qubit states}

\author{Berthold-Georg Englert,$^{*,\dag}$ %
Christian Kurtsiefer,$^\ddag$ %
and Harald Weinfurter$^{*,\ddag}$}

\address{$^*$Max-Planck-Institut f\"ur Quantenoptik, %
Hans-Kopfermannn-Stra\ss{}e 1, 85748 Garching, Germany\\%
$^\dag$Abteilung Quantenphysik, Universit\"at Ulm, %
Albert-Einstein-Allee 11, 89069 Ulm, Germany\\%
$^\ddag$Sektion Physik, Universit\"at M\"unchen, %
Schellingstra\ss{}e 4, 80799 M\"unchen, Germany} 

\date{To appear in Physical Review A, received 07 July 2000} 

\wideabs{
\maketitle
\begin{abstract}\hspace*{\absskip}%
Upon entangling a spatial binary alternative of a photon with its
polarization, one can use single photons to study arbitrary 2-qubit states.
Sending the photon through a Mach-Zehnder interferometer, equipped with sets of
wave plates that change the polarization, amounts to performing a unitary
transformation on the 2-qubit state.
We show that any desired unitary gate can be realized by a judicious choice of
the parameters of the set-up and discuss a number of applications. 
They include the diagnosis of an unknown 2-qubit state, an optical
Grover search, and the realization of a thought experiment invented by
Vaidman, Aharonov, and Albert.
\end{abstract}
}
\narrowtext

\section{Introduction}\label{sec:Intro}

Entangled qubits are central to most schemes that have been proposed for
quantum communication, quantum information processing, and
quantum cryptography (secure key distribution).
The basic unit consists of an entangled qubit pair.

Any binary quantum alternative can serve as a qubit and, therefore, different
degrees of freedom of one physical object can represent several qubits.
One could, for instance, encode some qubits in the motional degrees of freedom
of a trapped ion and other qubits in its internal degrees of freedom.
In our scheme, both qubits of an entangled pair are physically realized by a
single photon:
The photon's polarization is one qubit --- the ``polarization qubit'' --- and
the motional alternative of traveling to the right or to the left is the
second qubit --- the ``spatial qubit.''

It is our objective to present an optical model that facilitates experimental
studies of qubit pairs as realized by single photons.
Such single-photon 2-qubit states were used in a few recent experiments, 
including a variant of quantum teleportation \cite{BBdeMHP}, 
a remote state preparation \cite{MR-DBW},
demonstrations of simple quantum algorithms \cite{KMSW,Takeuchi}, 
a quantitative study of wave-particle duality \cite{SKE},
and a test of non-contextual hidden variable theories \cite{MWZ}.
Here we go beyond these special applications and consider arbitrary
manipulations of such states.

Studying qubit pairs extensively amounts to measuring observables of all
kinds.
The basic measurement is the detection of the photon in one of four standard
states given by combinations of
traveling to the right or left and polarized vertically or horizontally. 
This measurement is easily done, and experimental limitations are only due to
imperfections of optical elements (such as polarizing beam splitters) and the
efficiency of the single-photon detection.
More complicated observables are measured by first transforming the respective 
four eigenstates to the standard basis states, and then detecting those.
Accordingly, 
being able to perform arbitrary unitary transformations on 2-qubit states 
is tantamount to 
being able to measure arbitrary 2-qubit observables.

How this challenge is met, is shown in Sec.~\ref{sec:gates}, where we
present experimental set-ups that realize universal unitary gates --- for
either one of the qubits itself and for both of them jointly.
Then, in Sec.~\ref{sec:BasApp}, we turn to basic applications that include
controlled-not gates and the measurement of the Bell basis.
Advanced applications are discussed in Sec.~\ref{sec:AdvApp}: 
After dealing with the diagnosis of 2-qubit states and the Grover search, we
describe a proposal for a laboratory version of a thought experiment invented
by Vaidman, Aharonov, and Albert in 1987.
Indeed, their intriguing puzzle largely motivated the work reported here.
We close with a summary and outlook.
An appendix contains technical material of a more mathematical nature.

\begin{figure}[tb]
\begin{center}
\epsfig{file=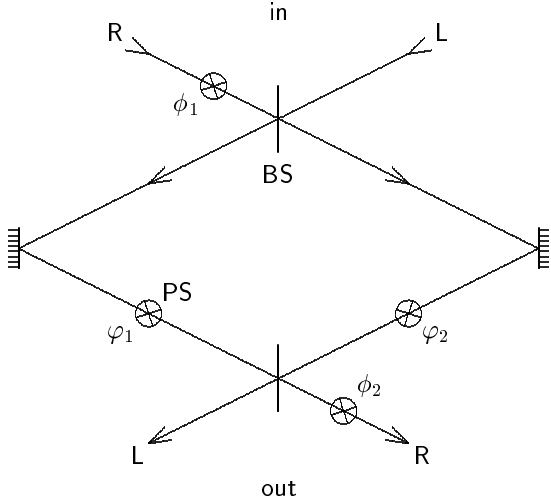}
\end{center}
\caption[Aa]{\label{fig:MZ}%
Mach-Zehnder set-up that realizes an arbitrary unitary gate for the spatial
$\R/\L$ qubit. There are symmetric beam splitters (BS's) at the entry and exit,
and four phase shifters (PS's) --- one each in the entry and exit $\R$ ports,
and two inside the interferometer. Additional PS's in the $\L$ ports
would be redundant; they could be introduced, either as a supplement or a
replacement of the PS's in the $\R$ ports, but there is no need for them.} 
\end{figure}

\section{Universal unitary gates}\label{sec:gates}
\subsection{Gates for the spatial qubit}\label{ssec:WWgate}
The spatial qubit consists of the binary alternative of moving to the right
($\R$) or to the left ($\L$), as indicated in the Mach-Zehnder geometry of
Fig.\ \ref{fig:MZ}.
As usual, we use analogs of Pauli's spin operators,
\begin{eqnarray}
&  \tau=\ketbra{\L}{\R}\,,\quad\tau\adj=\ketbra{\R}{\L}\,, &
\nonumber\\
&   \tau_1=\tau+\tau\adj\,,\quad\tau_2=i\tau-i\tau\adj\,,\quad
    \tau_3=\tau\adj\tau-\tau\tau\adj\,,&
\nonumber\\
& \idtau=\tau\adj\tau+\tau\tau\adj\,, &
  \label{eq:def-tau}
\end{eqnarray}
so that the unitary action of a symmetric beam splitter is given by
\begin{eqnarray}
  \UBS&=&\frac{1}{\sqrt{2}}\Bigl(\ketbra{\R}{\R}+\ketbra{\L}{\L}
                           +i\ketbra{\R}{\L}+i\ketbra{\L}{\R}\Bigr)
\nonumber\\
      &=&  \frac{1}{\sqrt{2}}\bigl(\idtau+i\tau_1\bigr)\,.
\label{eq:BS}
\end{eqnarray}
Likewise, the joint action of the mirrors inside the Mach-Zehnder set-up is
accounted for by the unitary operator
\begin{equation}\label{eq:mirr}
  \Umirr=-i\Bigl(\ketbra{\L}{\R}+\ketbra{\R}{\L}\Bigr)=-i\tau_1\,,
\end{equation}
where the inclusion of a phase factor $-i$ is a convenient convention
because it gives $\UBS\Umirr\UBS=\idtau$;
and phase shifters in the $\R$ and $\L$ branches amount to
\begin{eqnarray}
  \UR(\phi)&=&\ket{\R}\Exp{i\phi}\bra{\R}+\ketbra{\L}{\L}
            =\Exp{i\phi\tau\adj\tau}\,,
\nonumber\\
  \UL(\phi)&=&\ketbra{\R}{\R}+\ket{\L}\Exp{i\phi}\bra{\L} 
            =\Exp{i\phi\tau\tau\adj}\,.
  \label{eq:PS}
\end{eqnarray}
Putting these pieces together, one gets
\begin{equation}
\Bigl(\ket{\R},\ket{\L}\Bigr)\to\Bigl(\UMZ\ket{\R},\UMZ\ket{\L}\Bigr)
=\Bigl(\ket{\R},\ket{\L}\Bigr)\clUMZ
  \label{eq:MZ}
\end{equation}
for the whole Mach-Zehnder interferometer of Fig.\ \ref{fig:MZ}.
The unitary operator
\begin{eqnarray}
  \UMZ&=&\UR(\phi_2)\UBS\UR(\varphi_1)\UL(\varphi_2)\Umirr\UBS\UR(\phi_1)
\nonumber\\[1ex]
&=&\Exp{\frac{i}{2}(\phi_1+\phi_2+\varphi_1+\varphi_2)}
\nonumber\\ &&\times\Exp{\frac{i}{2}\phi_2\tau_3}
\Exp{\frac{i}{2}(\varphi_1-\varphi_2)\tau_2}\Exp{\frac{i}{2}\phi_1\tau_3}
  \label{eq:UMZ}
\end{eqnarray}
is represented by the numerical $2\times2$ matrix 
\begin{eqnarray}
\clUMZ&=&
\Exp{\frac{i}{2}(\varphi_1+\varphi_2)}\nonumber\\ &&\times
\left(\begin{array}{c@{\enskip\enskip}c}
\Exp{i(\phi_1+\phi_2)}\cos\frac{\varphi_1-\varphi_2}{2} &
\Exp{i\phi_2}\sin\frac{\varphi_1-\varphi_2}{2}  \\[1ex]
-\Exp{i\phi_1}\sin\frac{\varphi_1-\varphi_2}{2} &
\cos\frac{\varphi_1-\varphi_2}{2}
\end{array}\right)
  \label{eq:clMZ}
\end{eqnarray}
that multiplies the 2-component row $\bigl(\ket{\R},\ket{\L}\bigr)$ in
(\ref{eq:MZ}).
This matrix is slightly more general than the one in Eq.~(1) of
Ref.~\cite{RZBB}.  

The latter form in (\ref{eq:UMZ}), which is a parameterization 
in terms of three Eulerian angles 
$\phi_1$, $\varphi_1-\varphi_2 $, and $\phi_2$ combined with an over-all phase
factor, makes it obvious that any unitary operator for the $\R/\L$ qubit can
be realized by a Mach-Zehnder set-up of the kind shown in Fig.\ \ref{fig:MZ}. 
Note that $\UMZ=\idtau$ if $\phi_1=\phi_2=\varphi_1=\varphi_2=0$,
which is the reason for the conventional phase factor in (\ref{eq:mirr}).

\subsection{Polarization gates}\label{ssec:POLgate}
We regard vertical ($\v$) and horizontal ($\h$) polarization as the basic
alternatives of the polarization qubit, and the corresponding Pauli operators
are 
\begin{eqnarray}
&  \sigma=\ketbra{\h}{\v}\,,\quad\sigma\adj=\ketbra{\v}{\h}\,, &
\nonumber\\
&   \sigma_1=\sigma+\sigma\adj\,,\quad\sigma_2=i\sigma-i\sigma\adj\,,\quad
    \sigma_3=\sigma\adj\sigma-\sigma\sigma\adj\,,&
\nonumber\\
& \idsig=\sigma\adj\sigma+\sigma\sigma\adj\,. &
  \label{eq:def-sig}
\end{eqnarray}
The photon's polarization is manipulated with the aid of wave plates.
A quarter-wave plate (QWP), with its major axis at an angle $\theta$ to the
vertical direction, effects the transition
\begin{eqnarray}
  \Bigl(\ket{\v},\ket{\h}\Bigr)&\to&
   \Bigl(\UQWP(\theta)\ket{\v},\UQWP(\theta)\ket{\h}\Bigr)
\nonumber\\
&=&\Bigl(\ket{\v},\ket{\h}\Bigr)\clUQWP(\theta)\,,
  \label{eq:QWP}
\end{eqnarray}
where the unitary operator $\UQWP$ is given by
\begin{eqnarray}
  \UQWP(\theta)&=&\Exp{-i\theta\sigma_2}\Exp{-i\frac{\pi}{4}\sigma_3}
                  \Exp{i\theta\sigma_2}
\nonumber\\
&=&\Exp{-i\frac{\pi}{4}[\sigma_1\sin(2\theta)+\sigma_3\cos(2\theta)]}
\nonumber\\
&=&\frac{1}{\sqrt{2}}\bigl[\idsig-i\sigma_1\sin(2\theta)
                                            -i\sigma_3\cos(2\theta)\bigr]\,,
  \label{eq:UQWP}
\end{eqnarray}
and its $2\times2$ matrix representation reads
\begin{equation}  \label{eq:clUQWP}
\clUQWP(\theta)=\frac{1}{\sqrt{2}}\left(\begin{array}{c@{\quad}c}
1-i\cos(2\theta) & -i\sin(2\theta) \\
-i\sin(2\theta)  & 1+i\cos(2\theta)
\end{array}\right)\,.
\end{equation}
Likewise, the action of a half-wave plate (HWP) is accounted for by the
unitary operator
\begin{eqnarray}
  \UHWP(\theta)&=&\left[\UQWP(\theta)\right]^2=
\nonumber\\ &=& \Exp{-i\theta\sigma_2}\Exp{-i\frac{\pi}{2}\sigma_3}
                  \Exp{i\theta\sigma_2}
\nonumber\\ &=&-i\bigl[\sigma_1\sin(2\theta)+\sigma_3\cos(2\theta)\bigr]\,,
  \label{eq:HWP}
\end{eqnarray}
represented by the matrix
\begin{equation}
  \label{eq:clUHWP}
  \clUHWP(\theta)=\left[\clUQWP(\theta)\right]^2
=-i\left(\begin{array}{c@{\quad}c}
\cos(2\theta) &\sin(2\theta) \\
\sin(2\theta)  & -\cos(2\theta)
\end{array}\right)\,.
\end{equation}

\begin{figure}[tb]
\begin{center}
\epsfig{file=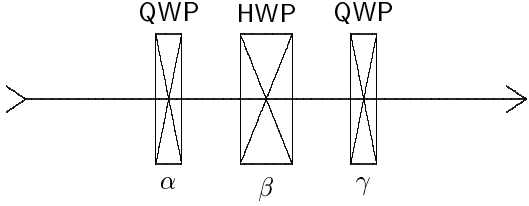}    
\end{center}
\caption[Aa]{\label{fig:3WPs}%
By sending a photon through a quarter-wave plate (QWP), then through a
half-wave plate (HWP), finally through another QWP, its polarization state can
be changed unitarily to any other one.}
\end{figure}

Particular polarization changes can be done with a single QWP, or a single
HWP, or with a QWP and a HWP in succession, and it is familiar \cite{WPs}
that the configuration of Fig.\ \ref{fig:3WPs}, where a HWP is
sandwiched by two QWP's, enables one to perform arbitrary changes of the
photon's polarization state. 
This is most easily seen by expressing the net unitary
operator in terms of three Eulerian angles,
\begin{eqnarray}
  \Upol&=&\UQWP(\gamma)\UHWP(\beta)\UQWP(\alpha)\nonumber\\
       &=&\Exp{-i(\gamma+\frac{3\pi}{4})\sigma_2}
        \Exp{i(\alpha-2\beta+\gamma)\sigma_3}
        \Exp{i(\alpha-\frac{\pi}{4})\sigma_2}\,.
  \label{eq:Upol}
\end{eqnarray}
We do not get an over-all phase factor here as there is in (\ref{eq:UMZ}), but
that does not matter.
For example, $\Upol=\idsig$ obtains for $\alpha=\beta\pm\pi/2=\gamma$ since
$\UQWP(\beta\pm\pi/2)=\bigl[\UQWP(\beta)\bigr]^{-1}$, and
$\alpha=\beta=\gamma$ gives  $\Upol=-\idsig$.
A polarization dependent phase shifter, that is
\begin{equation}
  \label{eq:polPS}
  \Upol=\ket{\v}\Exp{-i\vartheta}\bra{\v}
        +\ket{\h}\Exp{i\vartheta}\bra{\h}\,,
\end{equation}
is realized by the setting $\alpha=\gamma=\frac{1}{4}\pi$, 
$\beta=\frac{1}{2}\vartheta-\frac{1}{4}\pi$.

\subsection{Arbitrary 2-qubit gates}\label{ssec:2QBgate}
\begin{figure}[tb]
\begin{center}
\epsfig{file=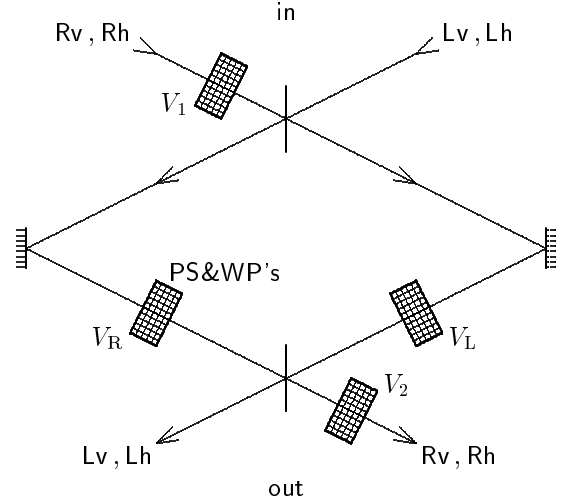}    
\end{center}
\caption[Aa]{\label{fig:2qbgate}%
Universal unitary gate for 2-qubit states.
In addition to the phase shifters (PS's) of Fig.\ \ref{fig:MZ}, there are now
wave plates (WP's) in the QWP/HWP/QWP combination of Fig.\ \ref{fig:3WPs}.
Each PS\&WP's set is specified by a phase (called $\phi_{1,2}$ or
$\varphi_{1,2}$ in Fig.\ \ref{fig:MZ}) and three angles $\alpha,\beta,\gamma$
that state the orientations of the WP's, as in Fig.\ \ref{fig:3WPs}.}
\end{figure}

Unitary gates $\UMZ$ and $\Upol$ for manipulations of the $\R/\L$ qubit and the
$\v/\h$ qubit individually are thus at hand.
We now combine them to construct universal gates that process arbitrary
2-qubit states unitarily.
This is achieved by a modification of the Mach-Zehnder set-up of
Fig.~\ref{fig:MZ}. 
In addition to the polarization-independent phase shifters already in place,
we let the photon pass through wave-plate combinations of the kind depicted in
Fig.~\ref{fig:3WPs}.
The entire set-up is then as shown in Fig.~\ref{fig:2qbgate}.

Where we had $\UR$ and $\UL$ in the product giving $\UMZ$ in (\ref{eq:UMZ}),
we now have corresponding factors in which the phase factors of (\ref{eq:PS})
are replaced by unitary operators that affect the polarization --- denoted by
$V_1$, $V_2$ for the entry and exit ports, and by $\VR$, $\VL$ inside the
interferometer. 
Each of them represents a phase shifter and a set of wave plates, and is
therefore of the form (\ref{eq:Upol}) multiplied by a phase factor.
Thus, the unitary operator $S$ associated with the 2-qubit gate of Fig.\
\ref{fig:2qbgate} is given by
\begin{eqnarray}
S&=&\left(\tau\adj\tau V_2+\tau\tau\adj\right)\UBS
\nonumber\\ &&\times
         \left(\tau\adj\tau\VR+\tau\tau\adj\VL\right)\Umirr
\nonumber\\ &&\times         
         \UBS\left(\tau\adj\tau V_1+\tau\tau\adj\right)\,,
  \label{eq:Ugate}
\end{eqnarray}
or
\begin{eqnarray}
S&=&\tau\adj\tau\SRR+\tau\tau\adj\SLL+\tau\SLR+\tau\adj\SRL
\nonumber\\[1ex]
&\repr&\left(\begin{array}{c@{\enskip}c}
\SRR & \SRL\\ \SLR & \SLL
\end{array}\right)_{\tau}\,,
  \label{eq:Ugate2}
\end{eqnarray}
where the $2\times2$ matrix refers to the spatial $\R/\L$ alternative, and the
entries of this matrix are
\begin{eqnarray}
\SRR&=&\frac{1}{2}V_2(\VR+\VL)V_1\,,
\nonumber\\
\SLL&=&\frac{1}{2}(\VR+\VL)\,,
\nonumber\\
\SRL&=&-\frac{i}{2}V_2(\VR-\VL)\,,
\nonumber\\
\SLR&=&\frac{i}{2}(\VR-\VL)V_1\,.
\label{eq:def-Ss}
\end{eqnarray}
The physical significance of these polarization operators is immediate:
$\SLR$, for instance, accounts for the polarization change associated with
photons entering the $\R$ port and leaving the $\L$ port.

\begin{figure}[tb]
\begin{center}
\epsfig{file=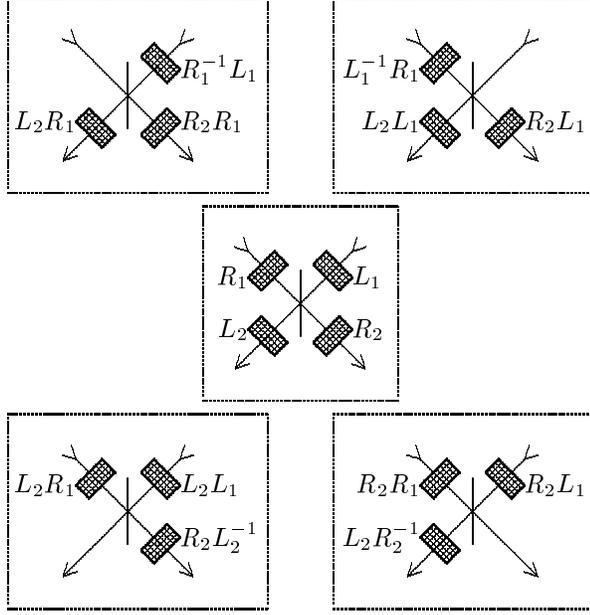}    
\end{center}
\caption[Aa]{\label{fig:EquivSetups}%
Equivalent set-ups involving a symmetric beam splitter and three or four sets
of phase shifter and wave plates. 
The central configuration has polarization-changing and phase-shifting
elements in both entry ports and both exit ports.
The two top configurations have one empty input port; 
the two bottom configurations have one empty output port.
With corresponding polarization gates, as indicated, each one of the five
set-ups represents the 2-qubit gate
$2^{-\frac{1}{2}}\bigl(\tau\adj\tau R_2R_1+\tau\tau\adj L_2L_1%
+i\tau L_2R_1+i\tau\adj R_2L_1\bigr)$.}
\end{figure}

There are no phase shifters or wave plates in the entry and exit $\L$ ports.
Indeed, one does not need them because the various combinations shown in
Fig.~\ref{fig:EquivSetups} are perfectly equivalent.  
Further configurations become possible when using polarizing beam splitters in
the Mach-Zehnder set-up.
Of course, when it comes to actual experimental realizations, one variant
could be more advantageous, for technical reasons, than the others
and then the freedom to choose freely among them is handy.
For the more theoretical purposes of the present discussion, however, we'll
confine ourselves to set-ups of the kind depicted in Fig.~\ref{fig:2qbgate}.  

The four operators in (\ref{eq:def-Ss}) need not be unitary themselves 
(and as a rule they aren't), 
but their form is much restricted by the unitary property of $S$, 
which implies the identities
\begin{eqnarray}
\SRR\adj\SRR\padj+\SLR\adj\SLR\padj&=&\idsig\,,\nonumber\\
\SRL\adj\SRL\padj+\SLL\adj\SLL\padj&=&\idsig\,,\nonumber\\
\SRR\adj\SRL\padj+\SLR\adj\SLL\padj&=&0\,,\nonumber\\
\SRL\adj\SRR\padj+\SLL\adj\SLR\padj&=&0\,,
  \label{eq:Sids}
\end{eqnarray}
the last two being adjoints of each other.
Since $V_1$, $V_2$, $\VR$, $\VL$ are unitary themselves, Eqs.\ (\ref{eq:Sids}) 
hold for the operators in (\ref{eq:def-Ss}) by construction. 

The reverse is also true: For any given unitary 2-qubit operator $S$ one can
find four unitary polarization operators $V_1$, $V_2$, $\VR$, $\VL$ such that
$S$ is of the form (\ref{eq:Ugate2}) with (\ref{eq:def-Ss}).
To prove this assertion, we must show that Eqs.\  (\ref{eq:def-Ss}) can be
solved for $V_1$, $V_2$, $\VR$, $\VL$ provided that the conditions
(\ref{eq:Sids}) are obeyed. 

A first technical step of this proof is given in the Appendix, where we
establish that $S\adj S=SS\adj=\idsig\idtau\equiv\id$ implies that the matrix 
entries of (\ref{eq:Ugate2}) are of the general form
\begin{eqnarray}
\SRR&=&\ket{\bpsi_1}\cos\vartheta\bra{\psi_1}
      +\ket{\bpsi_2}\cos\theta\bra{\psi_2}\,,\nonumber\\
\SLL&=&\ket{\bchi_1}\cos\vartheta\bra{\chi_1}
      +\ket{\bchi_2}\cos\theta\bra{\chi_2}\,,\nonumber\\
i\SRL&=&\ket{\bpsi_1}\sin\vartheta\bra{\chi_1}
       +\ket{\bpsi_2}\sin\theta\bra{\chi_2}\,,\nonumber\\
i\SLR&=&\ket{\bchi_1}\sin\vartheta\bra{\psi_1}
       +\ket{\bchi_2}\sin\theta\bra{\psi_2}\,,
  \label{eq:genSs}\label{EQ:APP}
\end{eqnarray}
where the kets and bras stand for particular sets of polarization states,
each set being orthonormal,
\begin{equation}
  \label{eq:ortnorm}
  \braket{\psi_j}{\psi_k}=
  \braket{\bpsi_j}{\bpsi_k}=
  \braket{\chi_j}{\chi_k}=
  \braket{\bchi_j}{\bchi_k}=\delta_{jk}\,,
\end{equation}
but with no other a priori relation among them.
Each set is specified by four parameters, two of them phases that do not enter
the basic projectors.
Since only states with the same subscript are paired in (\ref{eq:genSs}), 
six relative phases are relevant, so that two of the eight phases can be 
fixed by a convenient convention. 
In other words, 14 parameters are needed to specify the various ket-bra
products in (\ref{eq:genSs}).
Together with the values of $\vartheta$ and $\theta$, there is thus a total of
16 parameters, as there should be.

For given left-hand sides in (\ref{eq:genSs}), one determines the eigenvalues
and eigenstates of $\SRR\adj\SRR\padj$ to find $\vartheta$, $\theta$ and the 
$\psi$ states (with arbitrary phases).
The eigenstates of $\SRR\padj\SRR\adj$ then  supply the $\bpsi$ states with
well-defined phases relative to the $\psi$ states, and the eigenstates
of $\SLL\adj\SLL\padj$ and $\SLL\padj\SLL\adj$ yield the $\chi$ and $\bchi$
states, respectively. 

As soon as the ingredients of the right-hand sides of (\ref{eq:genSs}) are at
hand, one constructs the four $V$ operators in accordance with
\begin{eqnarray}
V_1&=&\ket{\chi_1}(\mp i)_1\bra{\psi_1}
      +\ket{\chi_2}(\mp i)_2\bra{\psi_2}\,,\nonumber\\
V_2&=&\ket{\bpsi_1}(\pm i)_1\bra{\bchi_1}
      +\ket{\bpsi_2}(\pm i)_2\bra{\bchi_2}\,,\nonumber\\
\VR&=&\ket{\bchi_1}\Exp{(\mp i)_1\vartheta}\bra{\chi_1}
      +\ket{\bchi_2}\Exp{(\mp i)_2\theta}\bra{\chi_2}\,,\nonumber\\
\VL&=&\ket{\bchi_1}\Exp{(\pm i)_1\vartheta}\bra{\chi_1}
      +\ket{\bchi_2}\Exp{(\pm i)_2\theta}\bra{\chi_2}\,,
  \label{eq:genVs}
\end{eqnarray}
where one must use consistently the upper or lower signs of $i$ in $(\ )_1$ and 
$(\ )_2$, but either one of the four possible sign choices will do.

\section{Basic applications}\label{sec:BasApp}
\subsection{Controlled-not gate}\label{ssec:CNOT}
As a first application, a warm-up problem, we consider controlled-not gates.
If the $\R/\L$ qubit controls the $\v/\h$ qubit, such a gate does nothing 
to the $\R$ input, but interchanges $\v\leftrightarrow\h$ on the $\L$ branch, 
\begin{eqnarray}
 && S_{{\rm cnot},\tau\to\sigma}\stdbaskets
\nonumber\\ &&\hspace*{8em}
=\Bigl(\ket{\R\v},\ket{\R\h},\ket{\L\h},\ket{\L\v}\Bigr)\,,
  \label{eq:cnot1}
\end{eqnarray}
where the subscript $\tau\to\sigma$ indicates which is the control qubit
($\tau$) and which the target qubit ($\sigma$).
Equivalently, we have
\begin{eqnarray}
 &  S_{{\rm cnot},\tau\to\sigma}=\tau\adj\tau\idsig+\tau\tau\adj\sigma_1\,,&
\nonumber\\
&\SRR=\idsig\,,\quad\SLL=\sigma_1\,,\quad\SRL=\SLR=0\,.&
  \label{eq:cnot2}
\end{eqnarray}
One possibility has the upper signs in (\ref{eq:genVs}), combined with
$\vartheta=\theta=0$ and
\begin{eqnarray}
&&\ket{\bchi_1}=\ket{\chi_2}=i\ket{\psi_1}=i\ket{\bpsi_1}=\ket{\v}\,,
\nonumber\\
&&\ket{\bchi_2}=\ket{\chi_1}=i\ket{\psi_2}=i\ket{\bpsi_2}=\ket{\h}\,,
  \label{eq:cnot2''}
\end{eqnarray}
so that
\begin{equation}
V_1=\VR=\VL=\sigma_1=i\UHWP(\pi/4)\,,\quad
V_2=\idsig\,,  
\label{eq:cnot2'} 
\end{equation}
which are easily realized with three HWP's and phase shifters that provide 
the factor of $i$.
We note that for a controlled-not gate, which interchanges
$\v\leftrightarrow\h$ on the $\R$ input but leaves the $\L$ input unchanged, a
single HWP for $V_1$ is sufficient.
No other polarization changing elements are needed ($V_2=\VR=\VL=\idsig$)
and thus the Mach-Zehnder interferometer isn't even necessary.
This is due to the specific configuration chosen in Fig.~\ref{fig:2qbgate}
where the $\L$ input is empty by convention
and, accordingly, for the gate defined by (\ref{eq:cnot1})
a single HWP (plus phase shifter) in the $\L$ input suffices, too.

If, however, the $\R/\L$ qubit is controlled by the $\v/\h$ qubit, 
\begin{eqnarray}
 &  S_{{\rm cnot},\sigma\to\tau}
=\idtau\sigma\adj\sigma+\tau_1\sigma\sigma\adj\,,&
\nonumber\\
&\SRR=\SLL=\sigma\adj\sigma\,,\quad\SRL=\SLR=\sigma\sigma\adj\,,&
  \label{eq:cnot3}
\end{eqnarray}
the Mach-Zehnder set-up is needed. 
Here one could use 
\begin{eqnarray}
&V_1=-i\idsig\,,\quad V_2=i\idsig\,,& \nonumber\\
&\VR=\idsig\,,\quad\VL=\sigma_3=i\UHWP(0)\,,
  \label{eq:cnot3'}
\end{eqnarray}
that is: phase shifters in the entry and exit $\R$ ports, nothing in the $\R$
branch of the interferometer, and a phase shifter plus a HWP in the $\L$ branch.

\subsection{Swapping gate}\label{ssec:swap}
The defining property of a swapping gate is its effect on a product state,
\begin{eqnarray}
&&  \bigl(\ket{\R}R+\ket{\L}L\bigr)\otimes\bigl(\ket{\v}v+\ket{\h}h\bigr)
\nonumber\\ & \longrightarrow & 
  \bigl(\ket{\R}v+\ket{\L}h\bigr)\otimes\bigl(\ket{\v}R+\ket{\h}L\bigr)\,,
  \label{eq:swap1}
\end{eqnarray}
where $R,L$ and $v,h$ are arbitrary probability amplitudes,
so that
\begin{equation}
  \label{eq:swap2}
  S_{\rm swap}\stdbaskets=
    \Bigl(\ket{\R\v},\ket{\L\v},\ket{\R\h},\ket{\L\h}\Bigr)\,,
\end{equation}
or
\begin{eqnarray}
& S_{\rm swap}=\frac{1}{2}\bigl(\id+\tau_1\sigma_1
                    +\tau_2\sigma_2+\tau_3\sigma_3\bigr)\,,&\nonumber\\
&  \SRR=\sigma\adj\sigma\,,\enskip\SLL=\sigma\sigma\adj\,,\quad
  \SRL=\sigma\,,\enskip\SLR=\sigma\adj\,.&
  \label{eq:swap3}
\end{eqnarray}
That $S_{\rm swap}$ interchanges the roles of the qubits is compactly stated
by
\begin{equation}
  \label{eq:swap3a}
  S_{\rm swap}\tau_k=\sigma_kS_{\rm swap}\quad\text{for $k=1,2,3$,}
\end{equation}
which can serve as an alternative definition.
The choice 
\begin{eqnarray}
V_1&=&-i\sigma_1=\UHWP(\pi/4)\,,\nonumber\\
V_2&=&i\sigma_1=\UHWP(-\pi/4)\,,\nonumber\\
\VR&=&\idsig\,,\quad  
\VL=-\sigma_3=-i\UHWP(0)\,,
\label{eq:swap4} 
\end{eqnarray}
(HWP's at the entry and exit, nothing in the $\R$ branch, phase shifter 
and HWP in the $\L$ branch) realizes the swapping gate.

\subsection{Walsh-Hadamard gate}\label{ssec:Had}
A Walsh-Hadamard gate turns the states of the standard basis into equal-weight
superpositions,
\begin{eqnarray}
&&S_{\rm WH}\stdbaskets\nonumber\\
&&\hspace*{1em}=\stdbaskets\frac{1}{2}
\left(\begin{array}{rrrr}
1&1&1&1\\
1&-1&1&-1\\
1&1&-1&-1\\
1&-1&-1&1
\end{array}\right)
  \label{eq:Hada1}
\end{eqnarray}
so that
\begin{eqnarray}
&S_{\rm WH}=\frac{1}{2}\bigl(\tau_1+\tau_3\bigr)\bigl(\sigma_1+\sigma_3\bigr)
\,,&\nonumber\\
\label{eq:Hada2}
&  \SRR=-\SLL=\SRL=\SLR=\frac{1}{2}\bigl(\sigma_1+\sigma_3\bigr)\,. &
\end{eqnarray}
A simple realization is specified by
\begin{eqnarray}
V_1&=&\idsig\,,\quad V_2=-\idsig\,,\nonumber\\[1ex]
\left.\begin{array}{r} \VR \\[0.5ex] \VL \end{array}\right\}&=&
-\frac{1\pm i}{2}\bigl(\sigma_1+\sigma_3\bigr)
=-i\Exp{\pm i\pi/4}\UHWP(\pi/8)\,.
  \label{eq:Hada3}
\end{eqnarray}
This choice needs nothing in the entry port, a phase shifter in the exit port, 
and HWP plus phase shifter in each arm of the interferometer.

\subsection{Bell basis measurement}\label{ssec:Bell}
\begin{figure}[tb]
\begin{center}
\epsfig{file=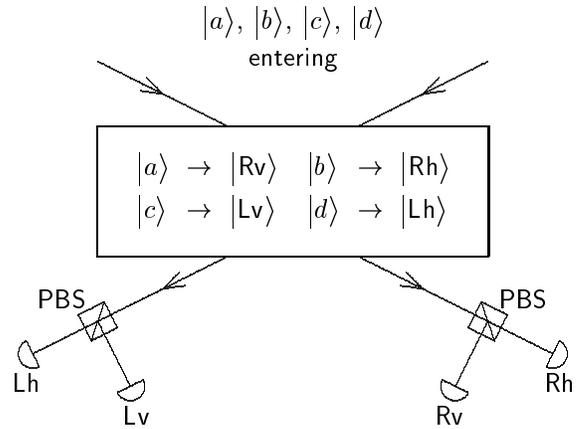}    
\end{center}
\caption[Aa]{\label{fig:BasisMeas}%
For a measurement of an arbitrary 2-qubit basis, consisting of the mutually 
orthogonal states $\ket{a}$, $\ket{b}$, $\ket{c}$, and $\ket{d}$, one first 
transforms it to the standard basis with the aid of an appropriate 2-qubit
gate.
The output is sent through polarizing beam splitters (PBS's) that reflect
vertically polarized photons and transmit horizontally polarized ones.
A click of either one of the four photon detectors (symbolized by semicircles) 
is indicative of the respective input state.}
\end{figure}

Another simple application is the measurement of the Bell basis, where we find
the 2-qubit photon in one of the four entangled superpositions
\begin{eqnarray}
\kBell{1}&=&2^{-\frac{1}{2}}\Bigl(\ket{\R\v}-\ket{\L\h}\Bigr)\,,\nonumber\\
\kBell{2}&=&2^{-\frac{1}{2}}\Bigl(\ket{\R\h}-\ket{\L\v}\Bigr)\,,\nonumber\\
\kBell{3}&=&2^{-\frac{1}{2}}\Bigl(\ket{\R\h}+\ket{\L\v}\Bigr)\,,\nonumber\\
\kBell{4}&=&2^{-\frac{1}{2}}\Bigl(\ket{\R\v}+\ket{\L\h}\Bigr)\,.
  \label{eq:Bell1}
\end{eqnarray}
Since one can detect the states of the standard basis --- viz.\  
$\ket{\R\v}$, $\ket{\R\h}$, $\ket{\L\v}$, and $\ket{\L\h}$ ---
with the aid of polarizing beam splitters (PBS's), see Fig.\
\ref{fig:BasisMeas}, all one needs is a 2-qubit gate that turns the Bell basis
into the standard one,
\begin{eqnarray}
 && S_{\rm Bell}\Bigl(\kBell{1},\kBell{2},\kBell{3},\kBell{4}\Bigr)
\nonumber\\&&\hspace*{6em}=\stdbaskets\,.
  \label{eq:Bell2}
\end{eqnarray}
Thus the ingredients
\begin{eqnarray}
  \label{eq:Bell3}
 &S_{\rm Bell}=2^{-\frac{1}{2}}\bigl(\idtau\idsig-i\tau_2\sigma_1\bigr)\,, & 
\nonumber\\
 &\SRR=\SLL=2^{-\frac{1}{2}}\idsig\,,\quad
  \SLR=-\SRL=2^{-\frac{1}{2}}\sigma_1&
\end{eqnarray}
are required. They are supplied by $V_1=V_2=\idsig$ in conjunction with 
\begin{eqnarray}
\VR&=&2^{-\frac{1}{2}}(\idsig-i\sigma_1)=\UQWP(\pi/4)\,,\nonumber\\
\VL&=&2^{-\frac{1}{2}}(\idsig+i\sigma_1)=\UQWP(-\pi/4)\,,
  \label{eq:Bell4}
\end{eqnarray}
for example, where one has just two QWP's inside the interferometer, one in
each branch, and nothing in the entry and exit ports. 

We note that an alternative way --- one of many --- of measuring the Bell
basis is stated by 
\begin{eqnarray}
 &&2^{-\frac{1}{2}} \bigl(\tau_1+\tau_3\bigr) S_{\rm swap} 
    S_{{\rm cnot},\sigma\to\tau}
    \Bigl(\kBell{4},\kBell{3},\kBell{1},-\kBell{2}\Bigr)
\nonumber\\&&\hspace*{6em}=\stdbaskets\,,
  \label{eq:Bell5}
\end{eqnarray}
where the permutation of the Bell states is irrelevant in the present context.
This measurement could be realized by a sequence of unitary transformations:
first a controlled-not gate (with $\v/\h$ controlling $\R/\L$), then a
swapping gate, finally a Walsh-Hadamard gate acting solely on the $\R/\L$ 
qubit; each of the three gates would require a Mach-Zehnder interferometer.
But rather than having three successive interferometers we can equivalently use
a single one, because \emph{any} unitary 2-qubit gate can be realized by the
set-up of Fig.~\ref{fig:2qbgate}, as shown in Sec.~\ref{ssec:2QBgate}.

\section{Advanced applications}\label{sec:AdvApp}
\subsection{State diagnosis}\label{ssec:diag}
As pointed out in the Introduction, we can measure any given 2-qubit
observable if we manage to detect its eigenstate basis, consisting of the
mutually orthogonal \mbox{2-qubit} states $\ket{a}$, $\ket{b}$, $\ket{c}$,
$\ket{d}$, say.
This is done, see Fig.\ \ref{fig:BasisMeas}, by mapping it onto the standard
basis. 
And, of course, it doesn't matter if this mapping involves additional phase 
factors. 
All one needs are transitions such as $\proj{a}\to\proj{\R\v}$.
In this context it is expedient to introduce two 2-qubit operators in
accordance with
\begin{eqnarray}
A&\equiv&\proj{a}+\proj{b}-\proj{c}-\proj{d}\,,\nonumber\\
B&\equiv&\proj{a}-\proj{b}+\proj{c}-\proj{d}\,,
  \label{eq:defAB}
\end{eqnarray}
so that $\ket{a}$, \ldots, $\ket{d}$ are the joint eigenkets of $A$ and $B$ 
with eigenvalues $A'=B'=+1$, \ldots, $A'=B'=-1$, respectively.  
The essential property of the unitary gate in Fig.\ \ref{fig:BasisMeas} is
then the mapping of $A$ and $B$ onto $\tau_3$ and $\sigma_3$,
\begin{eqnarray}
  \label{eq:ABmap}
  SA=\tau_3S\,,\quad SB=\sigma_3S\,.
\end{eqnarray}
For example, the operators $A=-\tau_1\sigma_1$ and $B=\tau_2\sigma_2$ 
are associated with the Bell basis (\ref{eq:Bell1}), and one verifies 
(\ref{eq:ABmap}) for $S_{\rm Bell}$ of (\ref{eq:Bell2}) easily.

Permutation of the basis states $\ket{a}$, \ldots, $\ket{d}$ have no effect on
the basis as a whole.
Therefore, one can interchange the roles of $A$ and $B$ in (\ref{eq:ABmap}),
or replace either one of them by their product $AB=BA$.
The respective gates are equivalent --- either one can be used to measure 
the basis in question --- but some may be simpler to set up than others.  
This is illustrated by the unitary transformation of (\ref{eq:Bell5}), which
corresponds to $A=\tau_1\sigma_1$ and 
$B=\tau_3\sigma_3=(-\tau_1\sigma_1)(\tau_2\sigma_2)$.

The statistical operator of a general 2-qubit state needs 15 real parameters
for its specification (see \cite{MetEng}, for example). 
The measurement of the probabilities associated with one 2-qubit basis
supplies three of the 15 parameters.
Accordingly, the full diagnosis of the 2-qubit state of interest
requires the measurement of at least five suitably chosen bases.

\begin{table*}[tb]
\caption[Aa]{\label{tbl:5sets}%
A minimal set of five $A,B$ pairs of 2-qubit observables. 
By measuring the corresponding 2-qubit bases, one determines all 15
parameters that specify the statistical operator of the given 2-qubit state.
The third column shows the unitary gates $S$ needed for the 
measurements, see Fig.~\ref{fig:BasisMeas}. 
The last four columns report possible choices for $V_1$, $V_2$, $\VR$, and $\VL$
that realize the respective $S$, see Fig.~\ref{fig:2qbgate}.
The $S$ of the first row is the Walsh-Hadamard gate of (\ref{eq:Hada2});
$\varepsilon$ is a stand-in for $\frac{1}{2}(1+i)$.
}
\begin{tabular}{ccccccc}
$A$ & $B$ & $S$ & $V_1$ & $V_2$ & $\VR$ & $\VL$
\\
\hline
$\tau_1$ & $\sigma_1$ & 
$\frac{1}{2}\bigl(\tau_1+\tau_3\bigr)\bigl(\sigma_1+\sigma_3\bigr)$ &
$\idsig$ & $-\idsig$ & 
$-\varepsilon(\sigma_1+\sigma_3)$ & $-\varepsilon^*(\sigma_1+\sigma_3)$
\\
$\tau_2$ & $\sigma_2$ &
$\frac{1}{2}\bigl(\idtau-i\tau_1\bigr)\bigl(\idsig-i\sigma_1\bigr)$ &
$i\idsig$ & $-i\idsig$ &
$\varepsilon(\idsig-i\sigma_1)$ & $\varepsilon^*(\idsig-i\sigma_1)$
\\
$\tau_3$ & $\sigma_3$ & $\id$ &
$\idsig$ & $\idsig$ & $\idsig$ & $\idsig$ 
\\
$\tau_1\sigma_2$ & $\tau_2\sigma_3$ &
$\frac{1}{2}\bigl(\id+\tau_2\idsig-i\idtau\sigma_2+i\tau_2\sigma_2\bigr)$ &
$\idsig$ & $\idsig$ &
$\idsig$ & $-i\sigma_2$
\\
$\tau_2\sigma_1$ & $\tau_3\sigma_2$ &
$\frac{1}{2}\bigl(\id-i\tau_2\idsig-i\tau_1\sigma_1-i\tau_3\sigma_1\bigr)$ &
$-i\idsig$ & $\sigma_1$ &
$\idsig$ & $i\sigma_1$
\end{tabular}
\end{table*}

A convenient set of such bases is reported in Table \ref{tbl:5sets}, where
each basis is characterized by its $A,B$ pair. 
These pairs identify five 2-qubit observables that are pairwise complementary
and thus optimal in the sense of Wootters and Field \cite{WooFie}.
In the terminology of Brukner and Zeilinger \cite{BruZei}, the five $A,B$'s are
``a complete set of five pairs of complementary propositions.'' 

Rather than using a minimal set of this kind, one could of course measure a
larger set of observables.
This was done by White \textit{et al.\/} \cite{WJEK}, who produced and studied
polarization-entangled photon pairs --- two qubits of the $\v/\h$ kind.
To our knowledge, theirs was the first experiment in which a complete
characterization of an entangled 2-qubit state was achieved.

\begin{figure}[!b]
\begin{center}
\epsfig{file=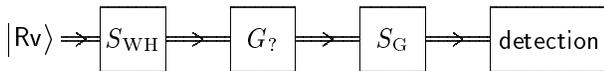}    
\end{center}
\caption[Aa]{\label{fig:Grover}%
Scheme of an optical implementation of Grover's search among four
possibilities. 
A photon in the 2-qubit state $\ket{\R\v}$ enters a Walsh-Hadamard
gate, then passes through the Grover gate, which performs either $G_1$,
$G_2$, $G_3$, or $G_4$. 
The photon is detected in one of the standard basis states, 
after being processed by $S_G$, and each of the four final states
corresponds uniquely to one of the four settings of the Grover gate.
Such an experiment was performed recently by Kwiat \textit{et al.\/}
\cite{KMSW}.} 
\end{figure}

\subsection{Grover search}\label{ssec:Grover}
In the present context of entangled 2-qubit states, Grover's problem 
\cite{Grov} amounts to the following; see Fig.~\ref{fig:Grover}.
Grover's gate applies either one of the four unitary operators
\begin{eqnarray}
G_1&=&\id-2\proj{\R\v}
=\frac{1}{2}\bigl(\id-\tau_3\idsig-\idtau\sigma_3-\tau_3\sigma_3\bigr)\,,
\nonumber\\
G_2&=&\id-2\proj{\R\h}
=\frac{1}{2}\bigl(\id-\tau_3\idsig+\idtau\sigma_3+\tau_3\sigma_3\bigr)\,,
\nonumber\\
G_3&=&\id-2\proj{\L\v}
=\frac{1}{2}\bigl(\id+\tau_3\idsig-\idtau\sigma_3+\tau_3\sigma_3\bigr)\,,
\nonumber\\
G_4&=&\id-2\proj{\L\h}
=\frac{1}{2}\bigl(\id+\tau_3\idsig+\idtau\sigma_3-\tau_3\sigma_3\bigr)
  \label{eq:Grov1}
\end{eqnarray}
to any 2-qubit state, and one has to find out which one 
is actually acting without using the gate more than once.

The solution consists of three steps.
First, we send a $\R\v$ photon through the Walsh-Hadamard gate of 
Sec.\ \ref{ssec:Had} to produce the superposition
\begin{equation}
  \label{eq:Grov2}
  \frac{1}{2}\Bigl(\ket{\R\v}+\ket{\R\h}
                +\ket{\L\v}+\ket{\L\h}\Bigr)\,.
\end{equation}
Second, this is used as input for Grover's gate, and the output is
\begin{equation}
  \label{eq:Grov3}
\begin{array}{r@{\enskip\mbox{for}\enskip}c}
\displaystyle
\frac{1}{2}\Bigl(-\ket{\R\v}+\ket{\R\h}
                +\ket{\L\v}+\ket{\L\h}\Bigr) & G_1\,, \\[1.5ex]
\displaystyle
\frac{1}{2}\Bigl(\ket{\R\v}-\ket{\R\h}
                +\ket{\L\v}+\ket{\L\h}\Bigr) & G_2\,, \\[1.5ex]
\displaystyle
\frac{1}{2}\Bigl(\ket{\R\v}+\ket{\R\h}
                -\ket{\L\v}+\ket{\L\h}\Bigr) & G_3\,, \\[1.5ex]
\displaystyle
\frac{1}{2}\Bigl(\ket{\R\v}+\ket{\R\h}
                +\ket{\L\v}-\ket{\L\h}\Bigr) & G_4\,. 
\end{array}
\end{equation}
Third, since these are four mutually orthogonal states, they can be mapped
onto the standard basis states, as in Fig.\ \ref{fig:BasisMeas}, here with the
unitary 2-qubit gate appropriate for $A=-\tau_3\sigma_1$ and 
$B=-\tau_1\sigma_3$ in (\ref{eq:ABmap}), viz.\
\begin{equation}
  \label{eq:Grov4}
  S_{\rm G}=\frac{1}{2}\bigl(\id-\tau_1\idsig-\idtau\sigma_1-\tau_1\sigma_1)\,.
\end{equation}
Thus, a click of the $\R\h$ detector, say, would tell us that $G_2$ was the
case. 

The choice
\begin{equation}
  \label{eq:Grov5}
  iV_1=-iV_2=-\VL=\idsig\,,\quad\VR=\sigma_1
\end{equation}
realizes $S_{\rm G}$ and thus offers a rather simple
single-photon implementation of Grover's search among four possibilities.

We note that Kwiat \textit{et al.\/} have already performed an
experiment of this kind \cite{KMSW}.  
These authors also discuss extensions to Grover searches among more than four
possibilities.

\subsection{Vaidman-Aharonov-Albert puzzle}\label{ssec:VAApuzzle}
Fitting to the present context, we rephrase the intriguing puzzle 
introduced by Vaidman, Aharonov, and Albert (VAA) in Ref.\ \cite{VAA} 
(and subsequently generalized by Ben-Menahem \cite{B-M} and 
Mermin \cite{Mermin}):
Chuck invites Doris to prepare two photons for him, photon~1 vertically
polarized and photon~2 in any polarization state she'd like.
He'll then perform a polarization measurement on photon~2, thereby measuring
either one of the three Pauli operators $\sigma_1$, $\sigma_2$, or $\sigma_3$,
without, however, telling Doris which one of the three complementary
measurements is actually done.
Since Chuck's measurement destroys photon~2, he promises to mimic an ideal
von Neumann measurement by turning the polarization of photon~1 from vertical
to the one detected for photon~2. 
Thereafter, Doris can measure any property of photon~1 allowed by
quantum mechanics.
Only after she did the measurement of her choosing, Chuck will tell Doris which
one of the three polarization measurements he had performed, and he challenges 
her to tell him then the outcome of his measurement.

Readers who don't know as yet how Doris can meet Chuck's challenge --- thereby
doing the seemingly impossible: ascertain the values of three mutually
complementary measurements --- should try to figure it out themselves before
reading on. 
There is a lesson here about the wonderful things entanglement can do for you.

\begin{figure}[!p]
\begin{center}
\epsfig{file=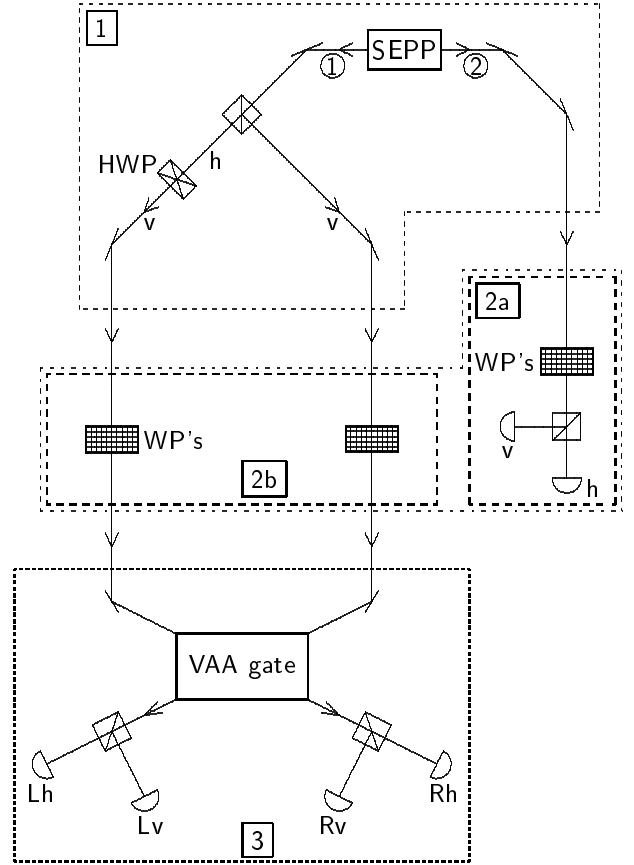}
\end{center}
\caption[Aa]{\label{fig:VAA}%
Proposed realization of the Vaidman-Aharonov-Al\-bert thought experiment of
Ref.~\cite{VAA}. 
It involves two photons (circled numbers) and consists of three stages 
(dashed boxes labeled by boxed-in numbers).
First stage: 
Doris prepares two photons for Chuck.
She uses polarization-entangled photons from a source of entangled photon 
pairs (SEPP). 
Photon~1 moves to the left and passes through a polarizing beam splitter. 
With a subsequent half-wave plate, Doris converts the transmitted, horizontally 
polarized, amplitude into vertical polarization. 
The photons are then no longer entangled in polarization. 
Instead, the polarization degree of freedom of photon~2 is now entangled with 
the spatial degree of freedom of photon~1. --- 
Second stage: 
(a)~Chuck measures the polarization of photon~2, 
either by distinguishing the linear polarizations $\v$ and $\h$, 
or the linear polarizations $\v\pm\h$, 
or the circular polarizations $\v\pm i\h$. 
Suitably set wave plates enable him to choose between the three 
complementary polarization measurements. 
(b)~Chuck then leaves a quantum record of his measurement result 
by changing the polarization of photon~1 from vertical 
to the just-detected polarization of photon~2. 
For this purpose he adjusts two sets of wave plates accordingly. ---
Third stage: 
With the aid of an appropriate unitary gate, such as the VAA gate 
specified by (\ref{eq:VAAgate}), Doris measures the VAA basis 
(\ref{eq:VAAbasis}) on photon~1. 
If Chuck then tells her which one of the three polarization measurements 
he did at the second stage, Doris can infer, with absolute certainty, 
the result he obtained.  
}
\end{figure}

Doris prepares the two photons in the entangled state
\begin{equation}
\label{eq:VAA1}
   2^{-\frac{1}{2}}\Bigl(\ket{(\R\v)_1\v_2}
                        +\ket{(\L\v)_1\h_2}\Bigr)\,.
\end{equation}
As shown in Fig.\ \ref{fig:VAA}, this is achieved by processing one photon of
a polarization-entangled pair emitted by a suitable source \cite{SEPP}
in the polarization state
\begin{equation}
  \label{eq:VAA2}
  2^{-\frac{1}{2}}\Bigl(\ket{\v_1\v_2}+\ket{\h_1\h_2}\Bigr)\,.
\end{equation}
Upon sending photon~1 through a polarizing beam splitter and rotating the
transmitted $\h$ polarization to $\v$, the polarization entanglement is turned
into an entanglement between the $\R/\L$ degree of freedom of photon~1 and the 
$\v/\h$ degree of freedom of photon~2, as described by the ket vector of
(\ref{eq:VAA1}). All of this happens during the first stage of the experiment
sketched in Fig.~\ref{fig:VAA}.

At the second stage, Chuck does one of the three polarization measurements.
If he measures $\sigma_1$, say, finding $\pm1$ leaves photon~1 in the state
\begin{equation}
  \label{eq:VAA3}
  2^{-\frac{1}{2}}\Bigl(\ket{\R\v}\pm\ket{\L\v}\Bigr)\,,
\end{equation}
and the subsequent change of its polarization from $\v$ to $\v\pm\h$ puts
photon~1 into 
\begin{equation}
  \label{eq:VAA4-1}
  \ket{{\sf 1}_{\pm}}\equiv
  \frac{1}{2}\Bigl(\ket{\R\v}\pm\ket{\R\h}\pm\ket{\L\v}+\ket{\L\h}\Bigr)\,.
\end{equation}
Likewise, if Chuck measures $\sigma_2$, photon~1 will emerge from the second
stage in one of the states
\begin{equation}
  \label{eq:VAA4-2}
  \ket{{\sf 2}_{\pm}}\equiv
  \frac{1}{2}\Bigl(\ket{\R\v}\pm i\ket{\R\h}\mp i\ket{\L\v}+\ket{\L\h}\Bigr)\,,
\end{equation}
and a measurement of $\sigma_3$ will produce
\begin{equation}
  \label{eq:VAA4-3}
  \ket{{\sf 3}_{+}}\equiv\ket{\R\v}\quad\text{or}\quad
  \ket{{\sf 3}_{-}}\equiv\ket{\L\h}\,.
\end{equation}
Note that these six states are simply related to the Bell states of
(\ref{eq:Bell1}),
\begin{eqnarray}
  \label{eq:VAA5}
  \ket{{\sf 1}_{\pm}}&=&2^{-\frac{1}{2}}\Bigl(\kBell{4}\pm\kBell{3}\Bigr)
\,,\nonumber\\
  \ket{{\sf 2}_{\pm}}&=&2^{-\frac{1}{2}}\Bigl(\kBell{4}\pm i\kBell{2}\Bigr)
\,,\nonumber\\
  \ket{{\sf 3}_{\pm}}&=&2^{-\frac{1}{2}}\Bigl(\kBell{4}\pm\kBell{1}\Bigr)\,.
\end{eqnarray}

At the third stage, Doris measures the VAA basis that consists of the states
defined by
\begin{equation}
  \label{eq:VAAbasis}
\left(\begin{array}{c}
\bVAA{1} \\[0.5ex] \bVAA{2} \\[0.5ex] \bVAA{3} \\[0.5ex] \bVAA{4} 
\end{array}\right)=
\frac{1}{2}\left(\begin{array}{rrrr}
 1 & -i &  1 & \phantom{-}1 \\[0.5ex]
 1 &  i & -1 & 1 \\[0.5ex]
-1 &  i &  1 & 1 \\[0.5ex]
-1 & -i & -1 & 1
\end{array}\right)
\left(\begin{array}{c}
\bBell{1} \\[0.5ex] \bBell{2} \\[0.5ex] \bBell{3} \\[0.5ex] \bBell{4} 
\end{array}\right)\,.
\end{equation}
The corresponding $A,B$ pair of observables and their product,
\begin{eqnarray}
 A&=&
\kBell{1}\bBell{4}+i\kBell{2}\bBell{3}-i\kBell{3}\bBell{2}+\kBell{4}\bBell{1}
\nonumber\\ 
&=&\frac{1}{2}\bigl(\tau_3\idsig+\idtau\sigma_3
                    +\tau_1\sigma_2-\tau_2\sigma_1\bigr)
\,,\nonumber\\[1ex]  
 B&=&
-i\kBell{1}\bBell{2}+i\kBell{2}\bBell{1}+\kBell{3}\bBell{4}+\kBell{4}\bBell{3}
\nonumber\\ 
&=&\frac{1}{2}\bigl(\tau_1\idsig+\idtau\sigma_1
                    -\tau_2\sigma_3+\tau_3\sigma_2\bigr)
\,,\nonumber\\[1ex] 
AB&=&
\kBell{1}\bBell{3}+i\kBell{2}\bBell{4}+\kBell{3}\bBell{1}-i\kBell{4}\bBell{2}
\nonumber\\
&=&\frac{1}{2}\bigl(-\tau_2\idsig+\idtau\sigma_2
                    +\tau_1\sigma_3+\tau_3\sigma_1\bigr)
=BA\,, 
 \label{eq:VAA-AB}
\end{eqnarray}
permute the states of the Bell basis.
The measurement of the VAA basis could, for example, employ a 2-qubit gate 
$S_{\rm VAA}$ that maps $A$ on $\tau_3$ and $B$ on $\sigma_3$, as in
(\ref{eq:ABmap}). 
One realization of this VAA gate is specified by
\begin{eqnarray}
  V_1&=& i\sigma_1= \UHWP(-\pi/4)\,,\nonumber\\
  V_2&=& \idsig\,, \nonumber\\
  \VR&=& \frac{1-i}{\sqrt{8}}\bigl(\idsig+i\sigma_1+i\sigma_2-i\sigma_3\bigr)
\nonumber\\ &=&\Exp{-i\pi/4}\UQWP(0)\UQWP(-\pi/4)\,, 
\nonumber\\
  \VL&=& \frac{1}{\sqrt{2}} \bigl(\idsig+i\sigma_2\bigr)
\nonumber\\ &=& \UQWP(\pi/4)\UQWP(0)\UQWP(-\pi/4)\,, 
\label{eq:VAAgate}  
\end{eqnarray}
which would need a HWP at the $\R$ entry, a phase shifter and two QWP's in
one arm, three QWP's in the other arm, and nothing at the exit.

\begin{table}[tb]
\caption[Aa]{\label{tbl:VAAprobs}%
Probabilities for Doris's measurement of the VAA basis 
(at the third stage of Fig.\ \ref{fig:VAA}) 
on the various states possibly prepared by Chuck (at the second stage).
}
\begin{tabular}{ccccccc}
Doris& \multicolumn{6}{c}{Chuck prepares} \\
finds 
 & $\ket{{\sf 1}_+}$ & $\ket{{\sf 1}_-}$
 & $\ket{{\sf 2}_+}$ & $\ket{{\sf 2}_-}$
 & $\ket{{\sf 3}_+}$ & $\ket{{\sf 3}_-}$ \\[1ex] \hline
$\bVAA{1}\rule{0pt}{2.5ex}$ 
&  $1/2$ & $0$   & $1/2$ & $0$ & $1/2$ & $0$ \\[1ex]
$\bVAA{2}$ &  $0$ & $1/2$   & $0$ & $1/2$ & $1/2$ & $0$ \\[1ex]
$\bVAA{3}$ &  $1/2$ & $0$   & $0$ & $1/2$ & $0$ & $1/2$ \\[1ex]
$\bVAA{4}$ &  $0$ & $1/2$   & $1/2$ & $0$ & $0$ & $1/2$ 
 \end{tabular}  
\end{table}

The probabilities listed in Table \ref{tbl:VAAprobs} are crucial in
understanding how Doris infers the result of Chuck's polarization measurement.
Suppose, for instance, that the $\L\v$ detector clicked, so that Doris found
photon~1 in state $\bVAA{3}$.
Then Chuck must have found $+1$ if he measured $\sigma_1$, and
$-1$ if he measured $\sigma_2$ or $\sigma_3$.
The VAA basis (\ref{eq:VAAbasis}) is, of course, chosen such that there are
enough entries `$0$' in Table \ref{tbl:VAAprobs}.

\section{Summary and outlook}\label{sec:sum}
We showed how one can manipulate, and thus study, entangled qubit pairs that
are physically represented by single photons.
One qubit is encoded in the polarization, the other in a spatial alternative
of the photon.
By purely optical means, one can perform arbitrary unitary transformations on
the qubit pair, so that any 2-qubit observable can be measured.
Potential applications include the complete diagnosis of the entangled
2-qubit state supplied by some source and the experimental realization of a
laboratory version of the Vaidman-Aharonov-Albert thought experiment.

The combined possibilities of performing any desired unitary transformation
and of measuring any observable of one's liking enables one to use qubit pairs
for other purposes as well.
In particular, any unitary 2-qubit gate is equivalent to a four-way
interferometer with certain relative phases between the four partial
amplitudes of certain strengths.
Therefore, a systematic quantitative study of four-way interferometers ---
that might ask questions concerning wave-particle duality, for example ---
could be done with single photons and 2-qubit gates of the kinds we discussed
above.  

Finally, we note that the set-up of Fig.\ \ref{fig:VAA} --- the optical
realization of the VAA thought experiment --- could be used for the purposes
of quantum cryptography.
Chuck, who would now control stages 1 and 2, sends single photons to Doris,
each photon in one of the six 2-qubit product states of Eqs.\ (\ref{eq:VAA5})
(which, incidentally, could be produced by different methods as well).
Doris, whose equipment would consist of the VAA gate and the photon detectors
in stage 3 of Fig.\ \ref{fig:VAA}, measures the VAA basis for each photon.
After receiving public word from Chuck which one of the three measurements he
performed at stage 2a, Doris infers his measurement results.
In this way, a random bit sequence is established that can serve as a
cryptographic key.
These matters are beyond the scope of the present paper and will be discussed
elsewhere \cite{BEKW}.

\section*{Acknowledgments}

BGE would like to thank Y. Aharonov for highly stimulating 
and most enjoyable discussions.
We are grateful for the insights gained in conversations with H.-J. Briegel.

\appendix

\section{Concerning Eqs.\ (\ref{EQ:APP})}
Equations (\ref{eq:Sids}) state $S\adj S=\id$ more explicitly.
Likewise $SS\adj=\id$ requires
\begin{eqnarray}
  \label{eq:app1}
  \SRR\padj\SRR\adj+\SRL\padj\SRL\adj&=&\idsig\,,\nonumber\\
  \SLR\padj\SLR\adj+\SLL\padj\SLL\adj&=&\idsig\,,\nonumber\\
  \SRR\padj\SLR\adj+\SRL\padj\SLL\adj&=&0\,,\nonumber\\
  \SLR\padj\SRR\adj+\SLL\padj\SRL\adj&=&0\,,
\end{eqnarray}
of which the last two are adjoints of each other.
We recall that, in a finite-dimensional Hilbert space as is the case here, the
selfadjoint products $X\adj X$ and $XX\adj$ are unitarily equivalent for any
operator $X$. 
When applied to $X=\SLR$, the first line in (\ref{eq:Sids}) and the second
line in (\ref{eq:app1}) imply that $\SRR\adj\SRR\padj$ and $\SLL\padj\SLL\adj$
are unitarily equivalent. 
Upon denoting their common eigenvalues by $\bigl(\cos\vartheta\bigr)^2$ and
$\bigl(\cos\theta\bigr)^2$, the eigenkets of $\SRR\adj\SRR\padj$ by 
$\ket{\psi_{1,2}}$ and those of $\SRR\padj\SRR\adj$ by $\ket{\bpsi_{1,2}}$,
the eigenkets of $\SLL\adj\SLL\padj$ by $\ket{\chi_{1,2}}$
and those of $\SLL\padj\SLL\adj$ by $\ket{\bchi_{1,2}}$, we then arrive at
the first two lines of (\ref{EQ:APP}). 
In doing so, some relative phases have been absorbed in the global phases of
the various kets and bras, but there remains the option to redefine them in
accordance with
\begin{equation}\label{eq:app2}
\begin{array}[b]{rcl@{\quad}rcl}
  \ket{\psi_k}&\to&\ket{\psi_k}\Exp{i\varphi_k}\,,&
  \ket{\bpsi_k}&\to&\ket{\bpsi_k}\Exp{i\varphi_k}\,,\\[1ex]
  \ket{\chi_k}&\to&\ket{\chi_k}\Exp{i\phi_k}\,,&
  \ket{\bchi_k}&\to&\ket{\bchi_k}\Exp{i\phi_k}\,,
\end{array}
\end{equation}
for $k=1,2$, without affecting the first two lines of (\ref{EQ:APP}). 

Next, the second line of (\ref{EQ:APP}) and the first line of (\ref{eq:app1})
tell us that
\begin{eqnarray}
  \label{eq:app3}
  \SRL\adj\SRL\padj&=&\idsig-\SLL\adj\SLL\padj\nonumber\\
&=&\ket{\chi_1}\bigl(\sin\vartheta\bigr)^2\bra{\chi_1}
+\ket{\chi_2}\bigl(\sin\theta\bigr)^2\bra{\chi_2}\,,\nonumber\\
  \SRL\padj\SRL\adj&=&\idsig-\SRR\padj\SRR\adj\nonumber\\
&=&\ket{\bpsi_1}\bigl(\sin\vartheta\bigr)^2\bra{\bpsi_1}
+\ket{\bpsi_2}\bigl(\sin\theta\bigr)^2\bra{\bpsi_2}\,,
\end{eqnarray}
with the consequence that $\SRL$ must be of the form
\begin{equation}
  \label{eq:app4}
  i\SRL=\ket{\bpsi_1}\Exp{-i\alpha}\sin\vartheta\bra{\chi_1}
+\ket{\bpsi_2}\Exp{-i\beta}\sin\theta\bra{\chi_2}\,,
\end{equation}
where $\alpha$ and $\beta$ are phases that are undetermined as yet.
Analogously, the first line of (\ref{EQ:APP}) and the second line of 
(\ref{eq:app1}) establish
\begin{equation}
  \label{eq:app5}
  i\SLR=\ket{\bchi_1}\Exp{i\alpha}\sin\vartheta\bra{\psi_1}
+\ket{\bchi_2}\Exp{i\beta}\sin\theta\bra{\psi_2}\,,
\end{equation}
where the phase factors are fixed by the third and fourth equations in 
(\ref{EQ:APP}) and (\ref{eq:app1}).

Now, the substitutions (\ref{eq:app2}) amount to
\begin{equation}
  \label{eq:app6}
  \alpha\to\alpha+\varphi_1-\phi_1\,,\quad\beta\to\beta+\varphi_2-\phi_2\,,
\end{equation}
in (\ref{eq:app4}) and (\ref{eq:app5}). Therefore, the phase factors 
$\Exp{\mp i\alpha}$ and $\Exp{\mp i\beta}$ can be removed by a suitable
redefinition of the kets and bras, and this turns (\ref{eq:app4}) and
(\ref{eq:app5}) into the last two lines of (\ref{EQ:APP}).

\end{document}